\documentclass{article}

\usepackage{arxiv}

\usepackage[utf8]{inputenc} 
\usepackage[T1]{fontenc}    
\usepackage{hyperref}       
\usepackage{url}            
\usepackage{booktabs}       
\usepackage{amsfonts}       
\usepackage{nicefrac}       
\usepackage{microtype}      
\usepackage{graphicx}
\usepackage[square, numbers]{natbib}
\usepackage{doi}
\usepackage{tabularx}
\usepackage{multirow}
\usepackage{color}
\usepackage{makecell}
\title{Text2Model: Generating dynamic chemical reactor models using large language models (LLMs)}

\author{ \href{https://orcid.org/0009-0007-6796-5359}       {\includegraphics[scale=0.06]{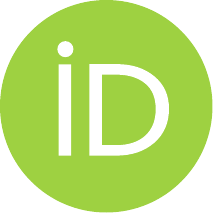}\hspace{1mm}Sophia Rupprecht}\\
	Process Intelligence Research Group\\
    Department of Chemical Engineering\\
    Delft University of Technology
\And
    Yassine Hounat\\
    Process Intelligence Research Group\\
    Department of Chemical Engineering\\
    Delft University of Technology
\And
    Monisha Kumar\\
    Process Intelligence Research Group\\
    Department of Chemical Engineering\\
    Delft University of Technology
\And
    \href{https://orcid.org/0009-0005-3475-9351}
    {\includegraphics[scale=0.06]{orcid.pdf}\hspace{1mm}Giacomo Lastrucci}\\
	Process Intelligence Research Group\\
    Department of Chemical Engineering\\
    Delft University of Technology
\And
	\href{https://orcid.org/0000-0001-8885-6847}{\includegraphics[scale=0.06]{orcid.pdf}\hspace{1mm}Artur M. Schweidtmann} \\
	Process Intelligence Research Group\\
    Department of Chemical Engineering\\
    Delft University of Technology\\
	\texttt{A.Schweidtmann@tudelft.nl}
}

\date{}

\renewcommand{\headeright}{} 
\renewcommand{\undertitle}{} 
\renewcommand{\shorttitle}{Text2Model}

\begin{document}
\maketitle

\begin{abstract}
As large language models have shown remarkable capabilities in conversing via natural language, the question arises as to how LLMs could potentially assist chemical engineers in research and industry with domain-specific tasks. We generate dynamic chemical reactor models in Modelica code format from textual descriptions as user input. We fine-tune Llama 3.1 8B Instruct on synthetically generated Modelica code for different reactor scenarios. We compare the performance of our fine-tuned model to the baseline Llama 3.1 8B Instruct model and GPT4o. We manually assess the models' predictions regarding the syntactic and semantic accuracy of the generated dynamic models. We find that considerable improvements are achieved by the fine-tuned model with respect to both the semantic and the syntactic accuracy of the Modelica models. However, the fine-tuned model lacks a satisfactory ability to generalize to unseen scenarios compared to GPT4o.
\end{abstract}

\section{Introduction}
Domain-specific languages for interaction with software tools or for data analysis require scientists and engineers to study the conventions and syntax of such languages \cite{Jacobs2024_LLMs4QuantumChemistrySim}. While software-user communication may often rely on a graphical user interface, such as AspenPlus or COMSOL, other tools require users to provide specialized code, e.g., MATLAB Simulink or Dymola. In addition to established programming languages for data analysis and visualization, such as Python or R, studying and applying a multitude of different domain-specific languages can be time- and labor-intensive.\\
Recent developments in using large language models (LLMs), particularly for code completion, code synthesis \cite{Nijkamp2022}, and code analysis and interpretation \cite{Wang2023_CodeT5Plus}, make such LLMs promising assistants for using tools such as modeling environments by converting textual information into domain-specific languages \cite{Jacobs2024_LLMs4QuantumChemistrySim}. State-of-the-art multilingual open models for code generation or understanding, such as CodeT5+ \cite{Wang2023_CodeT5Plus}, are literate in a range of programming languages in addition to natural language. Still, there are domain-specific languages in which pre-trained LLMs might not be proficient yet. The ability of state-of-the-art LLMs to converse in various programming languages motivates leveraging LLMs for synthesizing domain-specific languages analogously.\\
While there have been propositions that task-specific performance can be increased by few-shot learning \cite{Brown2020_LMsAreFewShotLearners}, fine-tuning is an established approach to tailor LLMs' responses to specific tasks \cite{Wei2021_FinetunedLMs_ZeroShotLearners}, especially for code synthesis \cite{Nijkamp2022}. Previous work by \citeauthor{Jacobs2024_LLMs4QuantumChemistrySim} has shown that it is possible to fine-tune GPT 3.5 to generate input files to ORCA, a programming package specific to the quantum chemistry domain \cite{Jacobs2024_LLMs4QuantumChemistrySim}. However, there is a lack of literature investigating the potential of LLMs to generate dynamic models for chemical reaction systems from descriptions in natural language.\\
We investigate how fine-tuning an open-source pre-trained LLM on dynamic reactor scenarios written in the domain-specific programming language Modelica affects the semantic and syntactic accuracy of such Modelica models generated by the fine-tuned model. We fine-tune Llama 3.1 8B Instruct to synthesize a full, simulatable Modelica model based on a textual description of a reactor scenario in the user's prompt. The fine-tuning dataset is generated synthetically from a set of templates of dynamic reactor scenarios. The supervised fine-tuning procedure uses low-rank adaptation~\cite{Hu2021_LoRA} as a parameter-efficient fine-tuning technique. We evaluate the results manually on samples randomly generated from the templated scenarios and on cases unseen during training. We compare the results of the fine-tuned model against the baseline Llama 3.1 8B Instruct and GPT4o.
\section{Methodology}\label{sec:Methods}
We develop (1) a methodology for the synthetic generation of the fine-tuning dataset (Section \ref{subsec:datagen}), (2) a supervised fine-tuning procedure (Section \ref{subsec:SFT}), (3) an evaluation procedure (Section \ref{subsec:Eval}), and (4) a set of evaluation metrics to assess the investigated models (Section \ref{subsec:metrics}).
\subsection{Data generation}\label{subsec:datagen}
The synthetic data used for training and evaluation consists of question-and-answer (Q\&A) pairs; the user provides a textual description of a dynamic tank reactor scenario in the question and an LLM is trained to provide a corresponding Modelica script as an answer. The training dataset is synthetically generated from templates of Q\&A pairs. The question contains a textual description of a reactor scenario in LaTeX syntax and the corresponding reference answer is a model in Modelica code presented as a string object. To provide further explanatory context, each line of Modelica code in the reference answers is appended with a descriptive comment of the defined parameter, the declared variable, or the expressed physics equation.\\
The reactor scenarios are designed to convey knowledge about modeling heuristics by independently varying the (a) mode of operation, (b) the reaction taking place, and (c) switching between ODE and DAE systems by removing the constant mixture density assumption. The mode of operations (a) are adiabatic isothermal, non-adiabatic isothermal, isothermal non-adiabatic, and non-isothermal non-adiabatic. For ODE systems (constant-density assumption), all four modes of operation are considered; for DAE systems, only the non-adiabatic non-isothermal mode is modeled. For the ODE scenarios, six reaction scenarios (b) are considered per mode of operation: Reversible and irreversible reactions between two, three, and four components, e.g., $\nu_A A+\nu_E E\rightarrow \nu_B B+\nu_C C$ or $\nu_A A\leftrightarrow \nu_B B+\nu_C C$. This yields 24 ODE scenarios in total by combining the variations of (a) and (b). For the single mode of operation (a) considered for DAE systems, two reaction scenarios are modeled: reversible and irreversible four-component reactions. This yields two DAE scenarios and, thus, 26 templated in total. Note that the stoichiometric numbers are assigned randomly for each template.\\
For each template, we randomly vary the units of measurement of given parameter values in the input question. The objective is to inspect whether performing direct conversions between non-SI and SI units with zero-shot LLM inference is attainable. Though Modelica provides conversion commands, e.g., from degrees Celsius to degrees Kelvin and vice versa, variables given in caloric energy units, for instance, must be converted to SI units manually.\\
We generate 988 Q\&A data points of which 790 are used for training and 198 are used for evaluation during training. Each of the 26 synthetic data templates is thus repeated 38 times with arbitrary values of given parameters. The parameter values are real numbers sampled from the range [10$^{-4}$,10$^5$ ] except for stoichiometric numbers, which are integer values between 1 and 15.
\subsection{Fine-tuning procedure}\label{subsec:SFT}
To enhance model performance in Modelica code generation, we fine-tune an existing LLM to adapt its model weights to better capture the nuances of this specific task. We use the model weights of the baseline Llama 3.1 8B Instruct model from Hugging Face~\cite{HF_LLama318BInstruct}.\\
The results presented in this work are obtained from fine-tuning using an NVIDIA~A100~80GB GPU with a batch size of one and a maximum sequence length of 2200 tokens. We re-use the same fast byte-pair-encoding tokenizer for tokenization as the baseline Llama 3.1 8B Instruct model. We train Llama 3.1 8B Instruct for eight epochs but evaluate the model obtained after four training epochs as indications of overfitting are observed after four epochs. We use linear learning rate scheduling with an initial learning rate of 0.0001. We apply a warmup ratio of 0.1, i.e., the initial learning rate of 0.0001 is reached after 632 steps.\\
We perform parameter-efficient finetuning by using the low-rank adaptation (LoRA) technique introduced by \citeauthor{Hu2021_LoRA}~\cite{Hu2021_LoRA}. We use a rank of eight and scale the accumulated gradient updates that are added to the frozen model weights by a factor of two. We thus effectively update 0.26~\% of the trainable 8.03 billion parameters. We do not apply quantization as initial studies have indicated shortcomings likely related to using quantization of model weights.
\subsection{Evaluation procedure}\label{subsec:Eval}
With the evaluation procedure presented hereafter, we aim to evaluate the effect of the fine-tuning. We evaluate the fine-tuned model, the baseline Llama 3.1 8B Instruct, and GPT4o on two evaluation datasets: The first dataset (1) tests whether the cases presented during training can be reproduced (Section~\ref{subsec:reproducability}: Examination of reproducibility of training scenarios); the second dataset (2) checks whether it is possible to extrapolate to reactor scenarios unseen during training (Section~\ref{subsec:Extrapolation}: Examination of generalization ability).\\
The first test dataset (1) (Section~\ref{subsec:reproducability}) contains entries that are randomly generated from each of the templates used for training with different parameter and variable values. This yields 26 samples in total. The second test dataset (2) (Section~\ref{subsec:Extrapolation}) includes four test cases: (a) a consecutive reaction, (b) two parallel reactions, and (c) a non-constant density system with five reaction participants (the reaction scenarios during training contain four components at the most). In the fourth case (d), we deliberately exclude parameter values that are necessary to fully specify the system. This investigates whether the missing parameter values are hallucinated in the models' responses or whether such user-generated errors are detected and addressed in the models' responses.\\
We perform inference on the baseline Llama 3.1 8B Instruct and the fine-tuned model using the same inference settings each. As a decoding strategy, we apply beam search with multinomial sampling using three beams and a temperature of 0.1. The maximum amount of tokens generated is limited to 4000. Inference is performed on the same machine as for fine-tuning, NVIDIA~A100~80GB.
\subsection{Evaluation metrics}\label{subsec:metrics}
We devise a customized set of evaluation metrics to quantify model performance. We manually evaluate each response given by an LLM (fine-tuned, baseline, GPT4o) by counting the number of occurrences of eight pre-defined error cases. These eight error cases examine general syntax errors, parameter and variable declaration errors, and errors in the system of equations. A detailed description of the eight error cases is provided below. A simplified example of a user prompt and a potential response by an LLM with errors is displayed in Figure 1. The Modelica code given in the response in Figure 1 contains illustrative examples of seven of the eight error cases considered in this study.\\
 For parameter and variable declaration errors, we distinguish between four subcases: unit conversion errors (1), declaration syntax errors (2), incorrect parameter values (3), and the usage of undefined variables (4). Errors in the equations section of the produced Modelica model are classified into incorrect equations (5), structural errors in the system of equations (6), and unit conflicts (7). General syntax errors (8) capture errors in the syntax of the model that would lead to compilation errors in the Dymola environment.\\
Unit conversion errors (1) arise from incorrect attempts to convert to SI units of measurement in cases where values given in the user's prompt are provided in non-SI units. Declaration syntax errors (2) are assigned in cases where Modelica syntax is not respected for parameter and variable declarations, e.g., in Figure 1, the initial value of cA0 is assigned with the word ‘initial’ instead of the Modelica convention ‘Concentration cA(start=cA0)’. Incorrect parameter values (3) originate from incorrectly transferred values given in the user's question to the output Modelica model. This error category does not punish unit conversion errors but examines fabricated values. Undefined variable errors (4) arise in cases where variables or parameters are used in the equations part of the Modelica model that have not been declared or defined before. Incorrect equations (5) are either physically incorrect or do not respect the constraints given in the users’ question, e.g., modeling an adiabatic system, although the user requested a non-adiabatic system. Structural errors in the system of equations (6) arise if the defined system of equations is structurally singular, i.e., the defined equations do not allow to determine all of the unknowns. Unit conflicts within the system of equations (7) occur when variables are defined in non-equivalent units of measurement, e.g., using Joules and Calories in the same equation, which do not cancel out. The result is that an unknown unit conversion factor distorts the lefthand-side value.
\begin{figure}[h!]
 \centering
    \includegraphics[width=\textwidth]{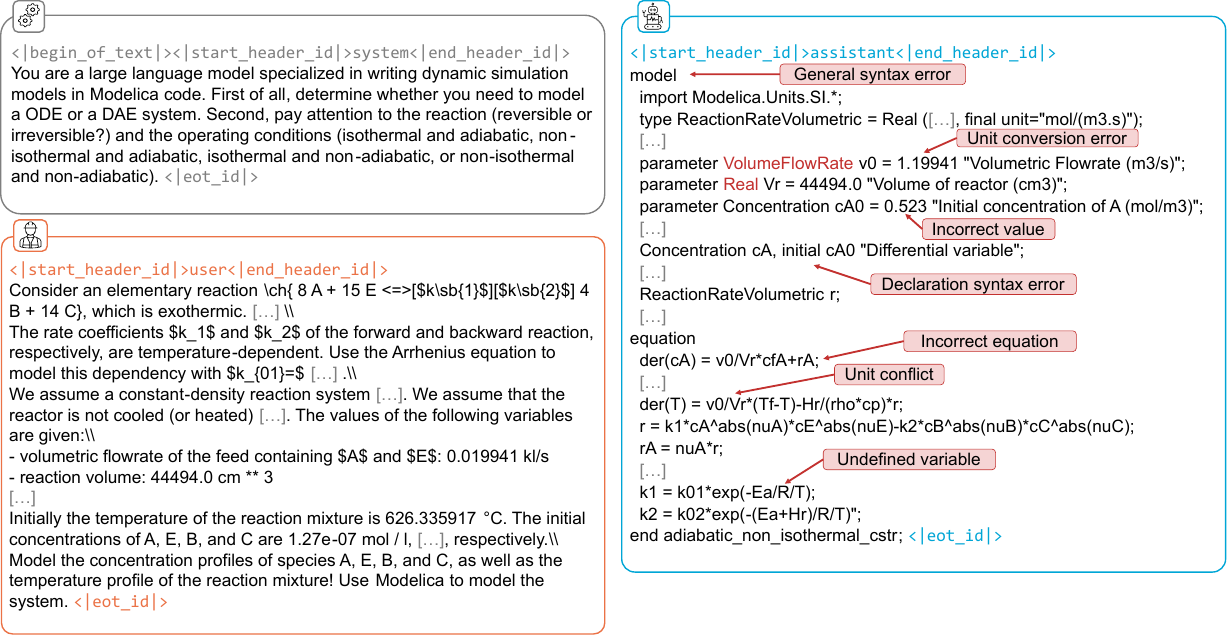}
 \caption{System message (top left), example user input (bottom left), and corresponding exemplary response (right). The example input and output are shortened for simplification. The output contains examples of seven error cases that are considered in this study.}
 \label{fig:ErrorSample}
\end{figure}
\section{Results and discussion}
We examine the results obtained for the two test datasets described in Section~\ref{subsec:Eval} separately (Section~\ref{subsec:reproducability} and \ref{subsec:Extrapolation}). In Section~\ref{subsec:reproducability}, we present the results for the first test dataset and examine the fine-tuned model’s ability to reproduce the reactor scenarios presented during training. In Section~\ref{subsec:Extrapolation}, we assess the second test dataset and inspect the fine-tuned model’s ability to extrapolate to reactor scenarios unseen during fine-tuning. We evaluate the fine-tuned Llama 3.1 8B model, the baseline Llama 3.1 8B model before fine-tuning, and GPT4o. For both test datasets, we use the same eight error cases described in Section~\ref{subsec:metrics} to assess the performance of all three models.\\
We observe that overall, considerable improvements are achieved by the fine-tuned model compared to the baseline Llama 3.1 8B Instruct for reactor scenarios seen during training (Section 3.1), as well as case studies containing elements unseen during training (Section 3.2). GPT4o shows remarkable performance of Modelica code generation and outperforms the fine-tuned model in cases unseen during training.
\subsection{Examination of reproducibility of training scenarios}\label{subsec:reproducability}
Compared to the baseline Llama 3.1 8B Instruct model, we observe that fine-tuning directly influences both the syntactical correctness of the produced Modelica model and the accuracy of the physics equations. Table~\ref{tab:reproduction} presents the absolute number of occurrences per error category for each examined model.
\begin{table}[h!]
\centering
\caption{Absolute amount of occurrences per error category per model. The test dataset comprises one randomly generated example for each of the 26 reactor scenario templates used for fine-tuning. The baseline model (Llama 3.1 8B Instruct) is fine-tuned and assessed on the same test dataset. GPT4o is assessed on the same test dataset as a benchmark.}
\begin{tabular}{c|cccc|ccc|c} 
        \toprule
        & \multicolumn{4}{c|}{\makecell{Parameter and variable declarations}}			
        & \multicolumn{3}{c|}{Physics equations} & \multirow{2}{*}{\makecell{General\\ syntax}} \\
            \cline{2-8}
        &\makecell{Unit\\error} & \makecell{Declaration\\syntax\\error} & \makecell{Incorrect\\values} & \makecell{Undefined\\variables} & \makecell{Incorrect\\equations} & \makecell{Structural\\error} & \makecell{Unit\\conflict} & \\
   \hline
        \makecell{Baseline\\(Llama 3.1 8B )} & 112 &	89 & 	\textbf{0} &	4 & 112 &	6 &	14 & 56 \\
        \hline
        \makecell{Fine-tuned\\(Llama 3.1 8B)} & \textbf{41} & \textbf{0} &	17 & 29 & \textbf{39} &	14	&\textbf{0} & 15\\
        \hline
        GPT4o & 50	& 27& 	1& 	\textbf{0}& 42&	\textbf{2}&	63& \textbf{4}\\
        \bottomrule
\end{tabular}   
\label{tab:reproduction}
\end{table}
The most prominent error categories of the baseline model are general syntax errors, unit conversion errors in parameter definitions, declaration syntax errors, and incorrect physical equations. On these four error categories, the fine-tuned model yields considerably fewer error cases: General syntax errors are reduced by 73.2~\% in the case of the fine-tuned model compared to the baseline model. Unit conversion errors are reduced by 63.4~\%, and declaration syntax errors are reduced to zero. In the error category of incorrect equations, the error cases produced by the fine-tuned model are reduced by 65.2~\% compared to the baseline model.\\
The fine-tuned model excels at handling unit conversions from non-SI units in the user's prompt in this first evaluation test set. The responses given by GPT4o contain conversion factors to transparently transform between units, while the fine-tuned model is trained towards directly converting between non-SI and SI units if needed. However, the manual unit conversion in GPT4o's responses is not done consistently in 63 cases, such that unit conflicts occur in the system of equations. In total, 113 combined error cases are produced by GPT4o in the context of unit conversion for parameter declaration and unit compatibility errors in equations compared to 41 in the case of the fine-tuned model, which is a reduction of 63.7~\%. The baseline Llama 3.1 8B Instruct model shows the highest number of error cases in the category unit conversion errors as, in contrast to GPT4o, unit conversions are done less systematically and are thus more error-prone. It should be considered to specifically instruct GPT4o and Llama 3.1 8B Instruct to convert to SI units directly in the system message provided during inference. The decrease of such unit conversion errors in the fine-tuned model gives rise to the assumption that it is reasonable to expect direct unit conversions. However, the fine-tuned model's robustness could benefit from training toward using manual conversion factors instead of direct conversion.\\
Most notably, the fine-tuned model consistently uses Modelica's built-in variable types, such as Concentration and Density. In contrast, most of the baseline and GPT4o's responses define parameters and variables as generic Real types. Using generic Real types for parameter definition and variable declarations does not leverage unit checks provided by the Dymola software when attempting to simulate a Modelica model.\\
Despite the improvements concerning the fine-tuned model's physical accuracy, we observe strong shortcomings in the reproducibility of DAE systems. The vast majority of incorrect physics equations produced by the fine-tuned model, 32 out of 39, occur in the two examples requesting DAE systems instead of ODE systems. In the case of the baseline model and GPT4o, physics errors related to DAE systems occur significantly less often, with eleven out of 112 and eleven out of 42, respectively. Furthermore, 13 out of 14 structural errors in the system of equations counted in the case of the fine-tuned model occur in the DAE test cases. As merely two of the 26 templated scenarios presented in the training data for fine-tuning contain DAE systems, the decreased performance in the cases of DAE systems during inference could be attributable to this imbalance in the training data. This assumption is supported by the observation that for ODE systems, the fine-tuned model indeed outperforms the baseline model and GPT4o.\\
A striking difference between the fine-tuned model and the baseline model and GPT4o is the occurrence of faulty parameter definitions with values that are most likely the result of hallucination. This is observed disproportionately often in the two test cases with DAE systems, with eight out of 17 total error cases. We thus hypothesize that hallucinations generally become increasingly frequent the more the test cases deviate from the majority of the training data.
\subsection{Examination of generalization ability}\label{subsec:Extrapolation}
The fine-tuned Llama 3.1 8B Instruct outperforms its pretrained counterpart on the four test cases that have not been part of the training data. We observe that GPT4o shows a higher physical accuracy of equations than the fine-tuned model. The absolute number of error occurrences per error category is displayed in Table \ref{tab:extrapolation}.
\begin{table}[h!]
\centering
\caption{Absolute amount of occurrences per error category per model. The test dataset comprises four case studies, each containing elements not portrayed in the fine-tuning dataset. The baseline model (Llama 3.1 8B Instruct) is fine-tuned and assessed on the same test dataset. GPT4o is assessed on the same test dataset as a benchmark.}
\begin{tabular}{c|cccc|ccc|c} 
        \toprule
        & \multicolumn{4}{c|}{\makecell{Parameter and variable declarations}}			
        & \multicolumn{3}{c|}{Physics equations} & \multirow{2}{*}{\makecell{General\\ syntax}} \\
            \cline{2-8}
        &\makecell{Unit\\error} & \makecell{Declaration\\syntax\\error} & \makecell{Incorrect\\values} & \makecell{Undefined\\variables} & \makecell{Incorrect\\equations} & \makecell{Structural\\error} & \makecell{Unit\\conflict} & \\
   \hline
        \makecell{Baseline\\(Llama 3.1 8B)} & 20 &	39 & 	3 &	\textbf{0} & 25 &	0 &	\textbf{0} & 9 \\
        \hline
        \makecell{Fine-tuned\\(Llama 3.1 8B)} & 15 & \textbf{0} &\textbf{0} & 2 & 14 & 0	&\textbf{0} & 1\\
        \hline
        GPT4o & \textbf{5}	& 3 & \textbf{0}& 	1& \textbf{8}&	0&	1& \textbf{0}\\
        \bottomrule
\end{tabular}
\label{tab:extrapolation}
\end{table}
Most strikingly, the performance of the fine-tuned model is worse than GPT4o in the categories of unit conversion errors in parameter definitions and the accuracy of physics equations. For instance, the reaction rate expressions of parallel and consecutive reactions (cases a and b, see Section~\ref{subsec:Eval}) are not formulated correctly in the responses of both the fine-tuned and the baseline models. In contrast, the responses provided by GPT4o contain correct rate expressions for both parallel and consecutive reactions. In the case of a requested non-constant density system (case c, see Section~\ref{subsec:Eval}), none of the models examined correctly formulate an algebraic equation for the reaction volume. In the case of left-out parameter values in the user's description (case d, see Section~\ref{subsec:Eval}), the responses of the fine-tuned model and the baseline model both hallucinate arbitrary values, GPT4o's response defines one parameter arbitrarily and does not define another at all.\\
We conclude that the fine-tuned model does not show satisfactory ability to extrapolate to scenarios not presented during training. Thus, it could be considered to re-train on a wider, more diverse fine-tuning dataset comprising a larger variety of Modelica models with extensive documentation alongside the reference answers. Furthermore, supporting external knowledge on, e.g., Modelica built-in variable types or textbooks on reactor modeling, could be provided to the LLM via, for instance, retrieval augmented generation (RAG)~\cite{Lewis2020_RAG}. Additionally, iterative feedback loops could be implemented by running the produced Modelica script in Dymola and re-routing failed attempts back for improvement.
\section{Conclusions}
We conclude that it is possible to leverage the potential of LLMs for generating dynamic models of chemical reactor systems from descriptions in natural language. We find that through fine-tuning Llama 3.1 8B Instruct, the syntactic and semantic accuracy of Modelica code generation can be improved compared to the baseline Llama 3.1 8B Instruct model. Compared to GPT4o, the baseline Llama 3.1 8B Instruct model and the fine-tuned counterpart do not show sufficient ability to extrapolate to scenarios unseen during training. We suggest improving model performance through three potential avenues: (1) expanding the fine-tuning dataset, (2) incorporating broader literature and Modelica documentation into the workflow by using RAG, and (3) iterative improvement by leveraging an interface to Dymola. The synthetic generation of a dataset comprising a diverse set of data (1) is labor- and time-intensive, and we also find no evidence of sufficient ability to extrapolate to scenarios not included in the training dataset. We thus suggest providing access to Modelica documentation and literature on reactor modeling (2) as a resource to increase the model's robustness in formulating physically accurate equations and using built-in Modelica variable types. Lastly, simulation of the generated Modelica script could be attempted in Dymola (3) and improved iteratively in case of failure.
\bibliographystyle{unsrtnat}
\bibliography{references}  






\end{document}